# ANALYSIS OF THE CARBONYL GROUP STRETCHING VIBRATIONS IN SOME STRUCTURAL FRAGMENTS OF POLY-3-HYDROXYBUTYRATE

G.A. Pitsevich[1*], E.N. Kozlovskaya[1], I.Yu. Doroshenko[2]

[1]Belarusian State University, Minsk, Belarus

[2]Taras Shevchenko National University of Kyiv, Kyiv, Ukraine

[*]*Correspondence author:* pitsevich@bsu.by

*The structure and the medium effects exerted on the spectral characteristics of the carbonyl group stretching vibrations in some structural fragments of poly-3-hydroxybutyrate have been analyzed. Calculations of the equilibrium configurations and IR spectra were carried out using the Gaussian program set in the approximation B3LYP/cc-pVDZ. It has been shown that typical bending of the poly-3-hydroxybutyrate chain is observed with an increase in the number of structural units. In order to explain the difference between the calculated and experimental frequencies of the C=O group stretching vibrations, the calculations of the potential energy curve associated with variations in the length of C=O bond and the subsequent numerical solution of a one-dimensional vibrational Schrödinger equation have been performed. The medium effects have been taken into account within the scope of a polarizable continuum model. Owing to the inclusion of the above-mentioned factors, which affect frequencies of the carbonyl groups stretching vibrations, correlation between the theoretical and experimental results has been improved significantly.*

*Key words: poly-3-hydroxybutyrate; hydrogen bond; potential energy surface, conformational analysis; polarizable continuum model.*

## 1. Introduction

Polyhydroxyalkanoates (PHA) are aliphatic polyesters synthesized by many microorganisms. On a number of properties they are similar to synthetic polymers. These polymers are characterized by thermoplasticity, optical activity, piezoelectric effect, biodegradability and biocompatibility. Combination of these properties, typical for PHA, makes their usage very promising in different fields. Based on the family of the PHA polymers, there is a possibility to obtain various materials with different mechanical properties. Poly-3-hydroxybutyrate (PHB) is the most famous and well studied member of the PHA.

As known, PHB consists of two parallel spiral structures, formed by the chain of hydrogen bonds. Hydrogen bond is formed between C=O group of one spiral structure and $CH_3$ group of other spiral structure [1]. Hydrogen bond C-H…O=C is bended [2]. An analysis of the registered IR spectra of the PHB [1, 3] indicates that most of the vibrational bands are associated both with the crystalline and amorphous states of the polymer. There are two bands in the region of stretching C=O vibrations – 1723 and 1740 cm$^{-1}$. Based on the temperature dependencies these bands are associated with vibrations of crystalline and amorphous carbonyl groups respectively. Besides experimental works, there are works devoted to quantum-chemical calculations of the PHB spectrum. The calculations confirm hydrogen bond formation and indicate that frequencies of C=O group stretching vibration in dimer are redshifted approximately by 20 cm$^{-1}$.

Necessity of the additional calculations for more accurate bands assignment and for determination of the nature of vibrational bands 1723 and 1740 cm$^{-1}$ is obvious. Besides, clarification of what exactly is meant by crystalline and amorphous PHB phases is needed. Calculations of the PHB IR spectrum considering influence of the solvent polarity are of interest, in particular – analysis of changes occurred in the spectrum with increasing of the medium polarity.

## 2. Calculation details

In order to analyze influence of the structure and media on spectral characteristics of the PHB carbonyl groups stretching vibrations, calculations of some its structural fragments were made using quantum-chemical set Gaussian [4]. Search of equilibrium configurations and calculations of IR spectra were made at B3LYP/cc-pVDZ [5-8] level of theory, which is acceptable for small amplitude vibrations in organic molecules according to [1, 9]. Generally, stretching vibrations of C=O bonds are small amplitude vibrations. However, the relatively high position of the first excited vibrational state above the potential energy curve minimum (approximately 3000 cm$^{-1}$ taking into account a zero-point energy) can cause anharmonicity effects [10, 11]. Considering --OC(CH$_3$)HCH$_2$C=O-- as a structural unit, the investigated structural fragments – model objects (MO) – can be represented as the ones that contain from two to six structural units with methyl and methoxy groups as the end groups. Equilibrium configurations of the CH$_3$[--OC(CH$_3$)HCH$_2$C=O--]$_n$CH$_3$ for n=2 and 5 are presented in Figure 1.

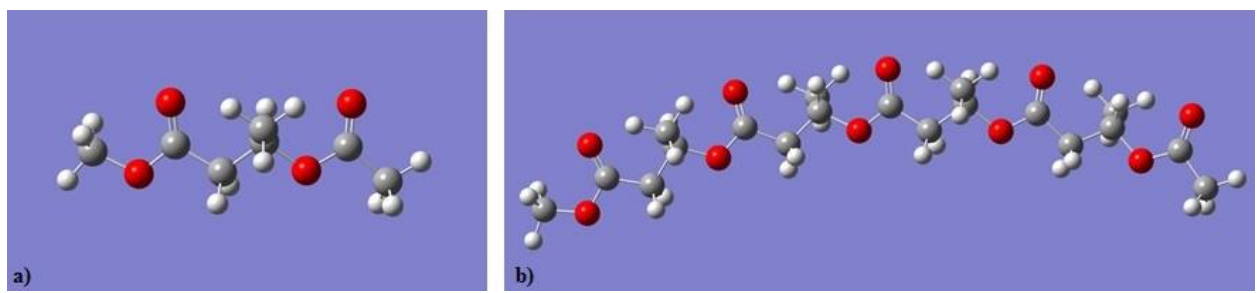

Figure 1. Equilibrium configurations of the PHB model fragments: a) n=2; b) n=5.

**3. Calculations of frequencies of the carbonyl groups stretching vibrations in model PHB fragments. Accounting of the anharmonicity and conformational activity effects**

Characteristic bending of the PHB chain is observed in the energetically preferable configuration with the increasing of structural units number (Fig. 1). This bending can lead to the spiral form of the PHB, which is observed experimentally [1]. The calculated frequencies and intensities of the C=O bond stretching vibrations in the investigated fragments are presented in Table 1.

Table 1. Calculated frequencies and intensities of $\nu_{C=O}$

| n=2 | | n=3 | | n=4 | | n=5 | | n=6 | |
|---|---|---|---|---|---|---|---|---|---|
| $\nu_{C=O}$ | I | $\nu_{C=O}$ | I | $\nu_{C=O}$ | I | $\nu_{C=O}$ | I | $\nu_{C=O}$ | I |
| 1819 | 353 | 1819 | 362 | 1819 | 359 | 1818 | 346 | 1819 | 357 |
| 1808 | 66 | 1810 | 205 | 1811 | 334 | 1811 | 422 | 1811 | 488 |
| | | 1799 | 39 | 1804 | 91 | 1805 | 186 | 1807 | 303 |
| | | | | 1797 | 14 | 1800 | 2 | 1803 | 6 |
| | | | | | | 1795 | 21 | 1798 | 7 |
| | | | | | | | | 1796 | 24 |

Analysis of the carbonyl group normal vibrations shows that they are collective vibrations. The most high-frequency vibration (1819 cm$^{-1}$) can be considered as an in-phase vibration. However, there is a trend in the shape of this vibration with n increasing: with small n values the amplitude of the C=O vibrations is approximately the same, while with large n values the amplitude begins to decrease in the direction from methyl end to the methoxy end. In the case of n=6, C=O bond that adjoins to the methoxy group does not participate in this vibration. One can suggest that with increasing of the PHB chain length, in-phase vibrations are localized on small structural fragments. Independence of the frequency of this vibration on the number of structural units indicates this indirectly (Table 1). Other vibrations of the carbonyl groups in one way or another are out-of-phase vibrations. With n increasing, more C=O groups (now starting from methoxy end) for 1811 cm$^{-1}$ vibration move in a phase. Their amplitude now decreases in the opposite of the case of 1819 cm$^{-1}$ vibration direction. The vibrations of the C=O bond, which adjoins methyl group, are out-of-phase ones. In general it is clear that the more C=O bonds vibrate in a phase and the less is the amplitude of the out-of-phase vibrations – the higher is the intensity of these vibrations.

This fact, in its turn, is the result of a co-direction of the C=O bonds in the energetically preferable configuration (all torsion angles of the carbon-oxygen skeleton are close to 180°). Due to this fact matching of the intensity ratio of in-phase and out-of-phase C=O bonds vibrations in the IR and Raman spectra should be expected, despite expectations noted in [1].

As follows from Table 1, the experimental values of absorption bands maxima are significantly lower than the computed ones. This can be due to a finiteness of the basis functions set. In order to examine this suggestion, geometry optimization and IR spectrum calculation of the MO with n=2 were made at the B3LYP/cc-pVTZ level of theory. Frequency values of in-phase and out-of-phase C=O vibrations appeared to be equal to 1800 and 1791 cm$^{-1}$. Comparing these data to the results of the calculations at the B3LYP/cc-pVDZ level of theory, one can assume that B3LYP/cc-pVDZ approximation systematically overestimates frequencies of the investigated vibrations by near 20 cm$^{-1}$. In order to estimate a contribution of the anharmonicity effects in the determination of the $\tilde{v}_{C=O}$ frequency, a one-dimensional potential energy surface (PES), associated with the length of the C=O bond adjoining to the methoxy group in the model object with n=2, was calculated at the B3LYP/cc-pVTZ level of theory. Bond length was changing in the interval -0.4–1.0 Å with step 0.1 Å. Other geometrical parameters remained the same. The calculated PES is presented in Figure 2.

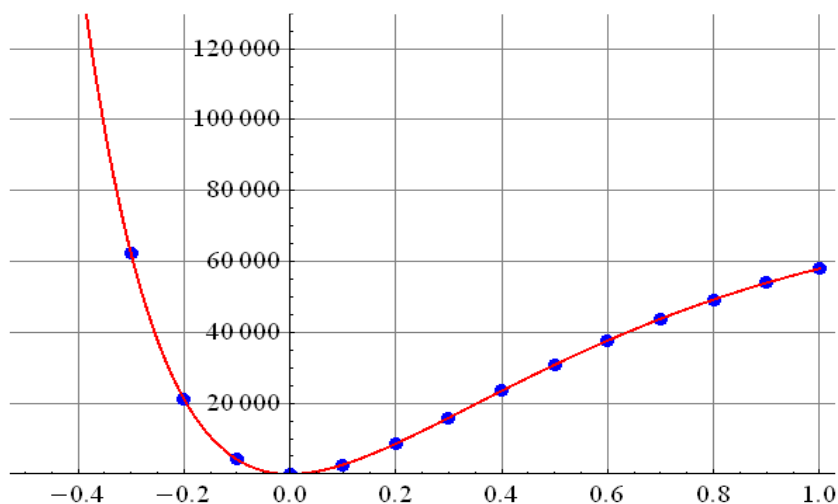

Figure 2. One-dimensional PES associated with C=O bond length variation

In order to obtain $\tilde{v}_{C=O}$, the following Schrödinger equation needs to be solved:

$$-\frac{\hbar^2}{2\mu_{C=O}l_0^2}\left[\frac{\partial^2}{\partial h^2}+\frac{2}{h_{C=O}^0+h}\frac{\partial}{\partial h}\right]\Psi+U(h)\Psi=E\Psi \quad (1)$$

Here $\mu_{C=O}$ - reduced mass of the oxygen and carbon atoms, $l_0$ – 1 Å, $\hbar$ – Planck constant, $h=q/l_0$; $q=l_{C=O}-l_{C=O}^0$; $h_{C=O}^0=l_{C=O}^0/l_0$; $l_{C=O}$ – length of the C=O bond, $l_{C=O}^0$ =1.2054Å – value of the equilibrium length of the C=O bond, $U(h)$ – potential energy, presented in Figure 2. Equation (1) was solved numerically by construction and following diagonalization of the Hamiltonian matrix. The method is described in detail in [12, 13]. Calculated values of the stationary energy levels and transition frequencies are presented in Table 2.

Table 2. Calculated values of the energy of $\nu_{C=O}$ vibrations and transition frequencies in MO n=2

| Vibrational quantum number | $E_n$, cm$^{-1}$ | $\tilde{\nu}_{C=O}$, cm$^{-1}$ | $2\tilde{\nu}_{C=O}$, cm$^{-1}$ | $3\tilde{\nu}_{C=O}$, cm$^{-1}$ | $4\tilde{\nu}_{C=O}$, cm$^{-1}$ |
|---|---|---|---|---|---|
| n=0 | 895.50 | | | | |
| n=1 | 2668.71 | 1773.21 | | | |
| n=2 | 4419.01 | | 3523.51 | | |
| n=3 | 6146.54 | | | 5251.04 | |
| n=4 | 7851.43 | | | | 6955.93 |

Using the well-known formula [14]:

$$\chi_{ii} = \frac{1}{2}\tilde{\nu}_i^{overton} - \tilde{\nu}_i^{fund};$$

anharmonicity constant value $\chi_{C=O/C=O}$ can be obtained for $\tilde{\nu}_{C=O}$ (-11.46 cm$^{-1}$). It indicates small anharmonicity of the analyzed vibrations. However, combining these two effects (finiteness of the basis set and anharmonicity effects) one can consider that harmonic calculations at the B3LYP/cc-pVDZ level of theory constantly overestimate $\tilde{\nu}_{C=O}$ frequency values by 50 cm$^{-1}$. In order to evaluate the influence of the conformational structure on the $\tilde{\nu}_{C=O}$ frequencies, configurations and IR spectra of some conformers of the model object n=4 were computed. An equilibrium configuration of the conformer I of the MO n=4 is presented in Figure 3.

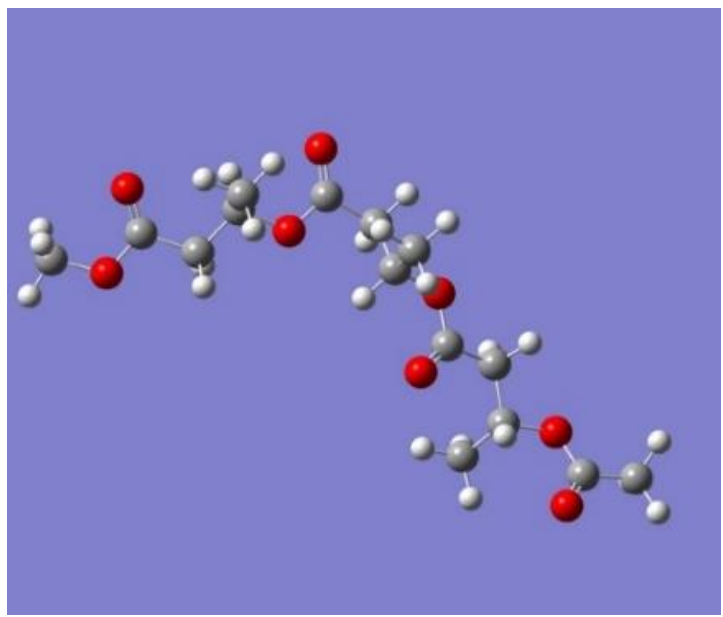

Figure 3. Equilibrium configuration of the conformer I of the MO n=4

Frequency values of the $\tilde{\nu}_{C=O}$ vibrations of three conformers of the MO n=4 are presented in Table 3.

Table 3. Calculated frequency values of the $\tilde{\nu}_{C=O}$ vibrations of the three conformers of the MO n=4

| n=4 Flat conformer | | n=4 Bent conformer I | | n=4 Bent conformer II | |
|---|---|---|---|---|---|
| $\nu_{C=O}$ | I | $\nu_{C=O}$ | I | $\nu_{C=O}$ | I |
| 1819 | 359 | 1819 | 176 | 1819 | 203 |
| 1811 | 334 | 1817 | 479 | 1817 | 462 |
| 1804 | 91 | 1807 | 60 | 1807 | 51 |
| 1797 | 14 | 1799 | 119 | 1799 | 115 |

Making no claim to commonality of the conclusions and remembering about finiteness of the objects with alternative configuration number, it should be noted that frequencies of the vibrations are almost insensitive to the conformational peculiarities of the MO, while the intensity of the out-of-phase vibrations can significantly increase with the transition from flat to bent configurations. The latter case is expected and explainable, because in flat configurations almost anti-parallel directed C=O bonds appear, and that leads to the increasing of the dipole momentum in the case of out-of-phase vibrations, and decreasing of the dipole momentum is the case of in-phase vibrations. Note as well that the intensity maximum shifts towards lower frequencies of the spectral interval 1819–1795 cm$^{-1}$ with n increasing. If this tendency will also take place in case of further increasing of n, then the agreement between the calculated and experimentally obtained frequencies will be obtained.

**4. Analysis of the crystalline and amorphous PHB phases. Accounting of the media polarity**

According to [1], the band 1723 cm$^{-1}$ is assigned to $\tilde{\nu}_{C=O}$ in a crystalline phase, and in an amorphous phase the band 1740 cm$^{-1}$ is assigned to $\tilde{\nu}_{C=O}$. Temperature dependence of the intensity of these bands speaks in favor of such assumption. Let`s assume, according to [1], that in the "crystalline" PHB phase spiral chains consisting of the energetically preferable flat configurations of the carbon-oxygen skeleton are bonded together with hydrogen bonds of C=O…H(CH$_2$) type, while in the "amorphous" phase spiral chains have limited length and are not bonded together with hydrogen bonds. It is clear that the effect of the conformational factor, presented in Table 3, would have the opposite to the experimentally registered temperature dependence of the intensity of the bands 1740 and 1723 cm$^{-1}$. With temperature increasing, the intensity of the 1723 cm$^{-1}$ band would also increase (see Table 3). Such tendency would have place if points, in which C----O skeleton deviates from the flat configuration, would often occur with temperature increasing. However, according to the data presented in [1], such points occur rarely even in the "amorphous" phase. So it should be considered that both bands are probably caused by in-phase C=O bands vibrations. At this the vibrational band at 1723 cm$^{-1}$ is associated with C=O vibrations, involved in hydrogen bonds formation between PHB spirals, and the 1740 cm$^{-1}$ band is associated with non-hydrogen bonded C=O bonds vibrations. Such pattern is generally consistent, because it correctly describes an absence of band shifts and presence of the intensity redistribution with temperature increasing. Moreover, according to this pattern, free C=O bonds will occur also in case of a strictly "crystalline" phase, and some of the groups involved in the hydrogen bonds formation will be present in a strictly "amorphous" phase, that correlates with experimentally registered case of the lack of total disappearance of one of the two absorption bands. However, there are facts that speak in favor of the different nature of the hydrogen bonds of the PHB spirals. If we consider the results presented in [1], where the distance 2.62 Å between O and H atoms is needed for the C=O…H(CH$_2$) bond formation, then, according to our calculations, intramolecular hydrogen bond has to appear between each side methyl group and one of the neighbor C=O groups, because in each of the investigated MO this distance do not exceed 2.62 Å. Since the number of formed interspiral hydrogen bonds per distance unit is considerably less than the number of intramolecular hydrogen bonds, then the transition from crystalline to amorphous phase is unlikely to be accompanied by such significant intensity redistribution of the bands 1740 and 1723 cm$^{-1}$. So, it can be assumed that interspiral hydrogen bonds have a different nature (for example, water molecules are acting as interspiral hydrogen bridge).

As is known from experimental data, when PHB is placed in KBr, the both absorption bands shift synchronously to the low-frequency region. One can conclude that there is no redistribution between the "crystalline" and "amorphous" phases. It can be suggested that there is a dry solution of the analyzed PHB in polar solvent. As is known, the value of the KBr dielectric constant is ε=5. In order to analyze the media polarity effect on the $\tilde{\nu}_{C=O}$ frequencies, equilibrium configurations and IR spectra of the previously analyzed conformers of the MO n=4 in media were calculated with $\varepsilon$=5 using PCM [15-17] at B3LYP/cc-pVDZ level of theory. The results of the calculations are presented in Table 4.

Table 4. Calculated $\tilde{\nu}_{C=O}$ frequency values for conformers of MO n=4 in media with ε=5.

| n=4 Flat conformer | | n=4 Bent conformer I | | n=4 Bent conformer II | |
| --- | --- | --- | --- | --- | --- |
| $\nu_{C=O}$ | I | $\nu_{C=O}$ | I | $\nu_{C=O}$ | I |
| 1795 | 660 | 1795 | 141 | 1795 | 124 |
| 1790 | 444 | 1794 | 909 | 1794 | 938 |
| 1783 | 109 | 1787 | 53 | 1786 | 49 |
| 1777 | 17 | 1780 | 178 | 1780 | 173 |

As is seen from comparison of the data in Tables 3 and 4, frequencies of all vibrations are redshifted approximately by 20-25 cm$^{-1}$. Making additional assumption that C=O groups, which participate in hydrogen bonds formation, are analogically redshifted, the experimentally observed situation with IR spectra changes can be explained as the result of increasing of the media polarity.

## 5. Conclusions

Geometry optimization and calculations of IR spectra of some structural fragments of PHB were made at the B3LYP/cc-pVDZ level of theory. The calculations have confirmed the experimentally registered bending of the chain. It was shown that the carbonyl group vibrations are collective vibrations. The most high-frequency vibration (1819 cm$^{-1}$) can be considered as the in-phase vibration. An assumption about localization of the in-phase vibrations on small structural fragments when the length of the PHB chain increases was made. Other carbonyl group vibrations are out-of-phase ones. An overestimation of the calculated frequencies of the C=O group vibrations in comparison to the experimental ones can also be associated with the basis set finiteness and with an anharmonicity effect. Using 1D PES calculations and numerical solution of the Schrödinger equation, it was estimated that these effects result in systematical overestimation of the frequencies of C=O vibrations approximately by 50 cm$^{-1}$ in harmonic calculations at the B3LYP/cc-pVDZ level of theory. This fact explains the discrepancies between the calculated and experimental frequency values. It was found that frequency values of C=O group stretching vibrations are almost insensitive to conformational peculiarities of PHB structural fragments, in oppose to the intensity, that can increase with the transition from the flat to the bent configuration. Taking into account the influence of the conformational factor, arguments are presented in favor of assumption about bands 1723 and 1740 cm$^{-1}$ being caused by in-phase C=O bands vibrations. The first band is assigned to carbonyl bonds vibrations, involved in hydrogen bond formation between PHB spirals, and the second band – to vibrations of the groups, which are not H-bonded. Besides, within the framework of calculations of equilibrium configurations and IR spectra of the PHB with taking into account the effect of the media polarity, we were able to adequately describe experimentally registered changes in IR PHB spectra.

**References.**